\documentclass[conference,compsoc]{IEEEtran}
\usepackage{amsmath}
\usepackage{amsfonts}
\usepackage{graphicx}
\usepackage{array}
\usepackage{xcolor}
\usepackage{epsfig}
\usepackage{stfloats}
\usepackage{amsthm}
\usepackage{booktabs, array, dcolumn}

\ifCLASSOPTIONcompsoc
  \usepackage[nocompress]{cite}
\else
  \usepackage{cite}
\fi
\usepackage{graphicx}
\usepackage{amsmath}
\usepackage{amsfonts}
\usepackage{graphicx}
\usepackage{array}
\usepackage{xcolor}
\usepackage{epsfig}

\ifCLASSINFOpdf
\else
\fi
\usepackage{subcaption}

\begin{document}
%
\title{BreathRNNet: Breathing Based  Authentication on 
\\Resource-Constrained  IoT Devices using RNNs}

\author{\IEEEauthorblockN{Jagmohan Chauhan\IEEEauthorrefmark{1}\IEEEauthorrefmark{2},
Suranga Seneviratne\IEEEauthorrefmark{1}\IEEEauthorrefmark{3},
Yining Hu\IEEEauthorrefmark{1}\IEEEauthorrefmark{2},
 Archan Misra\IEEEauthorrefmark{4},
Aruna Seneviratne\IEEEauthorrefmark{1}\IEEEauthorrefmark{2},
 Youngki Lee\IEEEauthorrefmark{4},
}
\IEEEauthorblockA{\IEEEauthorrefmark{1}School of EET (UNSW) , \IEEEauthorrefmark{2}Data61, Australia, \IEEEauthorrefmark{3} School of CSE (UNSW), \IEEEauthorrefmark{4} School of Information Systems (SMU)}
}


%


\maketitle

\begin{abstract}
Recurrent neural networks (RNNs) have shown promising results in audio and speech processing applications due to their strong capabilities in modelling sequential data. In many applications, RNNs tend to outperform conventional models based on \emph{GMM/UBMs} and \emph{i-vectors}. Increasing popularity of IoT devices makes a strong case for implementing RNN based inferences for applications such as acoustics based authentication, voice commands, and edge analytics for smart homes. Nonetheless, the feasibility and performance of RNN based inferences on resources-constrained IoT devices remain largely unexplored. In this paper, we investigate the feasibility of using RNNs for an end-to-end authentication system based on breathing acoustics. We evaluate the performance of RNN models on three types of devices; smartphone, smartwatch, and Raspberry Pi and show that unlike CNN models, RNN models can be easily ported onto resource-constrained devices without a significant loss in accuracy. 


\end{abstract}


%
\IEEEpeerreviewmaketitle

\section{Introduction}
\label{Sec:Introduction}


The variety of sensors embedded in smartphones, wearables, and other IoT devices are increasingly being used to support both (a) fine-grained monitoring of a user's activities and ambient context, and (b) richer forms of cyber-physical interaction (e.g., via gestures or natural language interfaces). Illustrative scenarios include the monitoring of a user's steps to estimate daily calorie expenditure, tracking of eating gestures to capture food intake and use of microphone-equipped devices at home for voice-based home automation control. As these personal and edge devices increasingly come in non-traditional form factors, as well as learn and store individual-specific information, it is important to develop novel and natural user authentication techniques. In recent work, we have introduced the \emph{BreathPrint}\cite{chauhan2017breathprint} system, which utilizes acoustic features of a user's breathing, captured by a commodity microphone, to support ubiquitous user authentication on mobile and IoT devices. 

 In this paper, we investigate whether RNN based deep learning models can be effectively used in resource-constrained devices for BreathPrint. The original work on BreathPrint~\cite{chauhan2017breathprint} was cloud based and used a conventional GMM based machine learning model with manually curated features. This work is motivated by two main objectives: {\bf{\emph{a) State-of-the-art speech recognition and speaker identification  methods are RNN based}}} and significantly outperform classical methods such as SVMs and GMM/HMMs especially in noisy environments~\cite{DeepEar}, thus it is important to evaluate the performance of BreathPrint using RNN models to boost the performance,  {\bf{\emph{b) For unobtrusive and ubiquitous authentication applications}}}, BreathPrint needs to be implemented in resource constrained devices and must be able to authenticate users without cloud access, thus it is important to obtain an idea about the accuracy vs. resource trade off for BreathPrint. To this end, in this paper we conduct a performance evaluation of an end-to-end RNN based BreathPrint authentication system on three representative hardware platforms:  IoT (Raspberry Pi 3), wearable (smartwatch) and mobile (smartphone). To the best of our knowledge this is the first work, which shows the feasibility, and performance of RNN based deep models for acoustics in limited resource footprint devices.

We make the following key contributions: 
\begin{itemize}
\item We present performance evaluation results of an end-to-end authentication based on LSTM (a variant of RNN) using breathing acoustics on three representative IoT devices: a smartphone, a smartwatch, and a Raspberry Pi.
\item We show that  RNN-based models for acoustic classification are smaller in size and lightweight than previously reported results with CNN-based models, and thus can be adopted to IoT devices. Specifically, an unoptimized RNN model is only 1.1 MB in size (for relevant breathing gestures) and can run on smartphones and smartwatches with approx. 100-200 ms and 700 m latencies, respectively.
\item We also show how a layer quantization-based model compression technique can help to reduce the memory footprint of RNN models (by a factor of 5, to approx 150-250 KB) without suffering any consequential drop in accuracy.
\end{itemize}

\section{Related Work}
\label{Sec:RelatedWork}
Motivated by the breakthroughs in training deep neural networks and the impressive performance gains achieved (deep networks have not only outperformed conventional machine learning models, but in several cases, even human experts), a variety of recent work has focused on the challenges (such as higher memory requirement or excessive computational latency) of executing deep learning models on resource-constrained devices. Table~\ref{tab:Works} summarizes some of these notable efforts and the techniques employed. Broadly speaking, these efforts utilize one or more of the following three approaches: (i) offloading  neural network processing to GPUs, which are more efficient in vectorized computations, (ii) reducing the time and memory requirements to load the fully-connected layers, and (iii) faster execution of the convolutional layers.

Early work by Lane et al.\cite{lane2015early} investigated the  performance characteristics, resource requirements and the execution bottlenecks for deep learning models (CNN and DNN) on mobile, wearable, and IOT devices, to support audio and vision based apps. Results indicated that  although smaller deep learning models work without issues on these devices, more complex CNN models such as AlexNet do not work well under the resource constraints.  To address this problem, Bhattacharya et al.~\cite{bhattacharya2016sparsification} proposed \emph{SparseSep}, which focuses primarily on finding a sparse representation of the fully connected layers and  on using separate filters to separate the convolutional kernels. These techniques reduce the number of parameters and convolutional operations required to execute a deep learning model, and can thus significantly reduce the computational and space complexity on resource-constrained devices. 

Several papers have focused on optimizing deep networks for audio and image sensing applications. DeepEar~\cite{DeepEar} is an audio sensing application for smartphones based on DNNs. DeepEar was implemented in the DSP of the smartphone, and imposed only 6\% additional overhead in daily energy consumption. DeepEye~\cite{mathurdeepeye} deployed CNNs on wearables for continuous vision applications. DeepEye avoids resource bottlenecks by using interleaving to orchestrate the execution of computation-heavy convolutional layers  with memory-heavy fully-connected layers. DeepEye also employs caching to load the fully connected layers faster and utilizes a singular-value decomposition (SVD) based layer factorization approach to compress the fully connected layer.  DeepMon \cite{huynh2017deepmon}  focused on reducing the processing latency  of convolutional layers, via multiple optimization techniques, for continuous vision applications. First, DeepMon employs a caching mechanism to take advantage of the likely significant similarities between consecutive images. Secondly, DeepMon perform faster matrix multiplication by model decomposition and unfolding. Model decomposition decomposes a convolutional layer into multiple smaller convolutional layers such that that the total computation of the decomposed layers is smaller than that of the original layer. Finally, DeepMon offloads convolutional layers to mobile GPUs for faster processing. More recently, MobiRNN \cite{cao2017mobirnn} applied GPU offloading to execute RNNs faster on smartphones, to support activity recognition tasks. 

As evident from the Table \ref{tab:Works}, most of the work to date has focused on CNNs and DNNs. For example, even audio analysis and speaker identification tasks have been performed using CNN and DNNs.  In general, CNNs are good at exploiting features defined on spatial data (e.g., images), whereas RNNs are more appropriate for identifying and using \emph{temporal} features, defined over data streams, such as audio or text.  Compared to CNN, RNN based models are also less complex as they deal with less complex data (images vs. text, audio) and do not use convolutional filters. Hence, compared to CNNs, RNN based models require comparatively lower computational power and memory~\cite{cao2017mobirnn, lane2015early}. 
 
As discussed earlier, \emph{BreathPrint} proposed a technique to authenticate users based on their breathing acoustics on mobile and IoT devices. \emph{BreathPrint} is based on the hypothesis that each individual's breathing pattern is unique; the proposed approach is also highly usable, as it merely requires the user to perform a small number of breathing gestures. Our interest in RNNs is thus driven by our belief that this uniqueness is manifested via temporal variations in a user's breathing pattern, and that RNNs are more capable in identifying and exploiting such temporal features. However, \emph{BreathPrint's} full potential can only be realized if the user identification can be performed locally (on the device), with minimal latency. Accordingly, in this work, we investigate the central question: \emph{``Can BreathPrint be practically realized, using an RNN-based model, on resource-constrained devices"?}


\begin{table}[t!]
\scriptsize
\centering
\caption{Deep Learning on Resource-Constrained Devices} \vspace{-3mm}
\begin{tabular}{p{1cm}|p{1.2cm}|p{1.2cm}|p{1.5cm}|p{1.5cm}}\specialrule{.12em}{1em}{0em}
{\bf Name} &  {\bf Type of DL} &  {\bf Architecture} &  {\bf Application} &  {\bf Techniques} \\ \hline

SparseSep &  CNN, DNN  & Multiple Layers & Image Classification, Speaker Identification,  Scene Analysis & Sparsification and Separation  \\ \hline
DeepEar &  DNN  & Five Layers & Audio Sensing & NA  \\ \hline
Deepeye &  CNN  & Multiple Layers & Continuous Vision& Interleaving, Caching and Compression  \\ \hline
DeepMon &  CNN  & Sixteen Layers & Continuous Vision& GPU Offloading, Caching and Decomposition  \\ \hline
MobiRNN &  RNN  & Two Layers & Activity Recognition   & GPU Offloading \\ \hline

\specialrule{.12em}{0em}{0em}
\end{tabular} \vspace{-4mm}
\label{tab:Works}
\end{table}

\section{Experimental Setup}
\label{Sec:Experiment}


To evaluate the feasibility of RNN-driven breathing-based authentication, we utilize the breathing acoustics dataset collected in our previous work~\cite{chauhan2017breathprint}. The dataset consists of acoustics samples of three breathing gestures; \emph{deep breathing}, \emph{normal breathing}, and \emph{sniffing} (two quick inhalations) of 10 users collected over three sessions. For each gesture the dataset contains 30, 30 and 10 samples collected on first day (session 1), fourth day (session 2) and seventh/eighth day (session 3)  respectively. In this paper, we focus only on two breathing gestures \emph{deep} and \emph{sniff}, as our earlier investigations revealed that those two perform better in authentication applications compared to \emph{normal breathing}. For further details on the dataset, please refer to our original paper~\cite{chauhan2017breathprint} that details the data collection process and the dataset.

For each user, we selected the first 50 samples for model training and tuning purpose. As deep learning requires larger sample sizes for training, we applied two commonly used data augmentation techniques to increase the number of data samples. In particular, we used a combination of the frequency wrapping technique~\cite{islam2017soundsifter}, and the amplitude scaling method~\cite{nguyen2017swallownet}. Each sample was scaled 10 times along the time axis and the amplitude by selecting two separate values from a uniform distribution; $\sim$$\mathcal{U}$$(0.8,1.2)$. Overall, we obtain a 11-fold boost in the number of training examples (550 training samples per participant), consisting of both original samples and their augmented versions. The remaining 10 ``original" samples each, from session 2 and session 3, were kept intact for testing.







We performed experimental evaluation on using four devices, belonging to three distinct \emph{types}, listed in Table~\ref{tab:Devices}. The three types of devices include two smartphones (mobile), a smartwatch (wearable) and a Raspberry Pi (IoT). All the devices run different variant of Android based OSes, and representative of popular commercial mobile, wearable and embedded platforms. \\


%

\begin{table}[t!]
\scriptsize
\centering
\caption{Hardware Configuration of the Used Devices} \vspace{-3mm}
\begin{tabular}{p{1.5cm}|p{2cm}|p{1.1cm}|p{0.9cm}|p{1cm}}\specialrule{.12em}{1em}{0em}
{\bf Device} &  {\bf OS} &  {\bf CPU} &  {\bf GPU} &  {\bf Memory} \\ \hline

Nexus 5 &  Android  6.0& 2.26 GHz \newline Quad-core & Adreno 330 & 2$\;$GB  \\ \hline
Pixel &  Android  7.0& 2.15 GHz \newline Quad-core & Adreno 530 & 4$\;$GB  \\ \hline
LG G Watch R & Android Wear 2.0& 1.2 GHz  \newline  Quad-core& Adreno 305 & 512$\;$MB\\ \hline
Raspberry Pi  3  &  Android Things 5.0 & 1.2 GHz  \newline  Quad-core& VideoCore IV  & 1$\;$GB   \\ \hline

\specialrule{.12em}{0em}{0em}
\end{tabular} \vspace{-4mm}
\label{tab:Devices}
\end{table}

\section{Methodology}
\label{Sec:Methodology}

Our overall goal is \emph{user authentication} -- i.e., deciding whether a sample belongs to one of $N$ pre-registered possible users. This is effectively a problem of \emph{closed set user identification} that can be mapped as a multi-class classification problem (where a single class represents one user). Our approach is to do some pre-processing of the original acoustics signal so that it can be fed into an RNN  model. For a comparative baseline that uses a \emph{shallow} classifier, we also train a SVM model. We discuss the details of these steps below.



\subsection{Feature Extraction}
We divided each audio file into 10$\;$ms \emph{non-overlapping frames} with Hamming window based smoothing.  For each frame we calculated 96 MFCC features (32 MFCC, 32 Delta MFCC, and 32 Double Delta MFCC) using JSTK (Java Speech ToolKit).  Then we used \emph{windowing} to combine these frames so that temporal information between the frames is  retained. For sniff and deep breathing gestures, we tried window sizes of length=\{20, 25, 30, 35\} and \{200, 250, 300,  350\} respectively. There are two factors to consider when selecting a suitable window size. On the one hand, each window must be large enough to retain a significant part of a breathing gesture. On the other hand, the duration of a single breath varies significantly across users; consequently, if a larger window size is selected, some testing samples may be missed from users with relatively short breathing durations. To balance these considerations,
we ended up with window sizes \{20, 25, 30\} for sniff and \{200, 250\} for deep breathing gestures, respectively. To further augment the training dataset we created overlapping windows for a given breathing sample. We chose three \emph{overlap} sizes 90\%, 70\%, and 50\% of the window size. For each window size and overlap value pair, we trained the classifiers as discussed below.

\subsection{Training and Testing Datasets}
We used the first 50 samples from session~1 and session~2 during the training phase. More specifically we created the windows from these samples and randomly shuffle them and used 80\% of the windows as the {\bf \emph{training set}} and the rest of the 20\% as the {\bf \emph{validation set}} to tune hyper parameters.  We refer to the windows created from the rest of the 10 audio samples in session 2 as the  {\bf \emph{intra set}} and windows created from 10 audio samples from session 3 as {\bf \emph{inter set}}.

 \subsection{RNN Model}
 
\begin{figure}[h!]
\centering
\epsfig{file=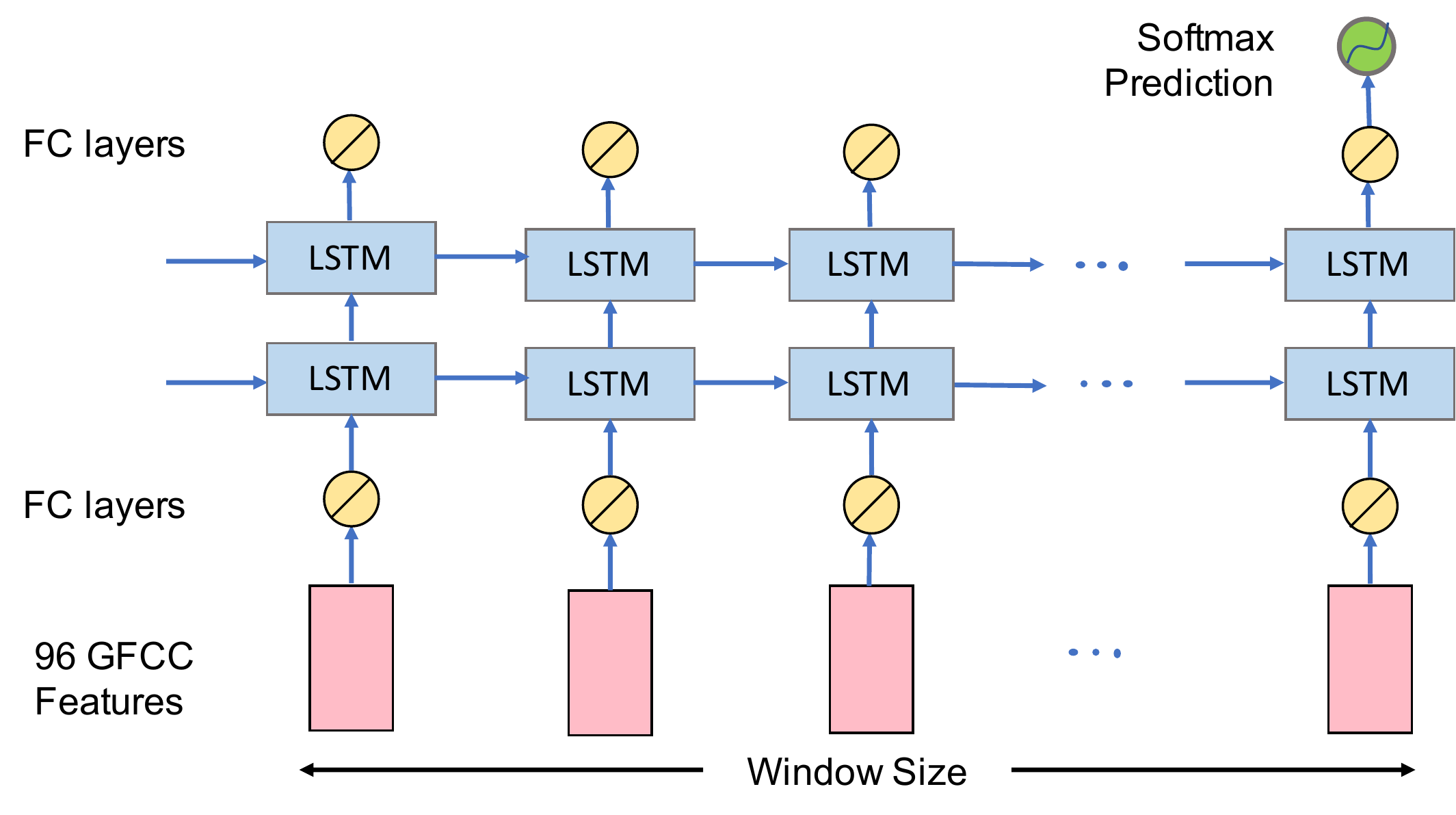,scale=0.37}
\caption{RNN Architecture}
\label{fig:RNN}
\end{figure}

 We used an RNN architecture that is similar to Hammerla et al.~\cite{Hammerla:2016}; this architecture is illustrated in Figure~\ref{fig:RNN}. The hidden unit size of the LSTM units was 128 and we used two LSTM layers. We implement the model using \emph{Tensorflow} ~\cite{abadi2016tensorflow}. We used 32 as batch size and trained the network over 500 iterations. As a baseline classifier we also trained a multi-class SVM classifier with linear kernel using \emph{libSVM}.

 \subsection{Model Selection}
\begin{figure}[t!p]
         \begin{subfigure}[b]{0.25\textwidth}
                 \centering
                 \includegraphics[width=0.98\textwidth]{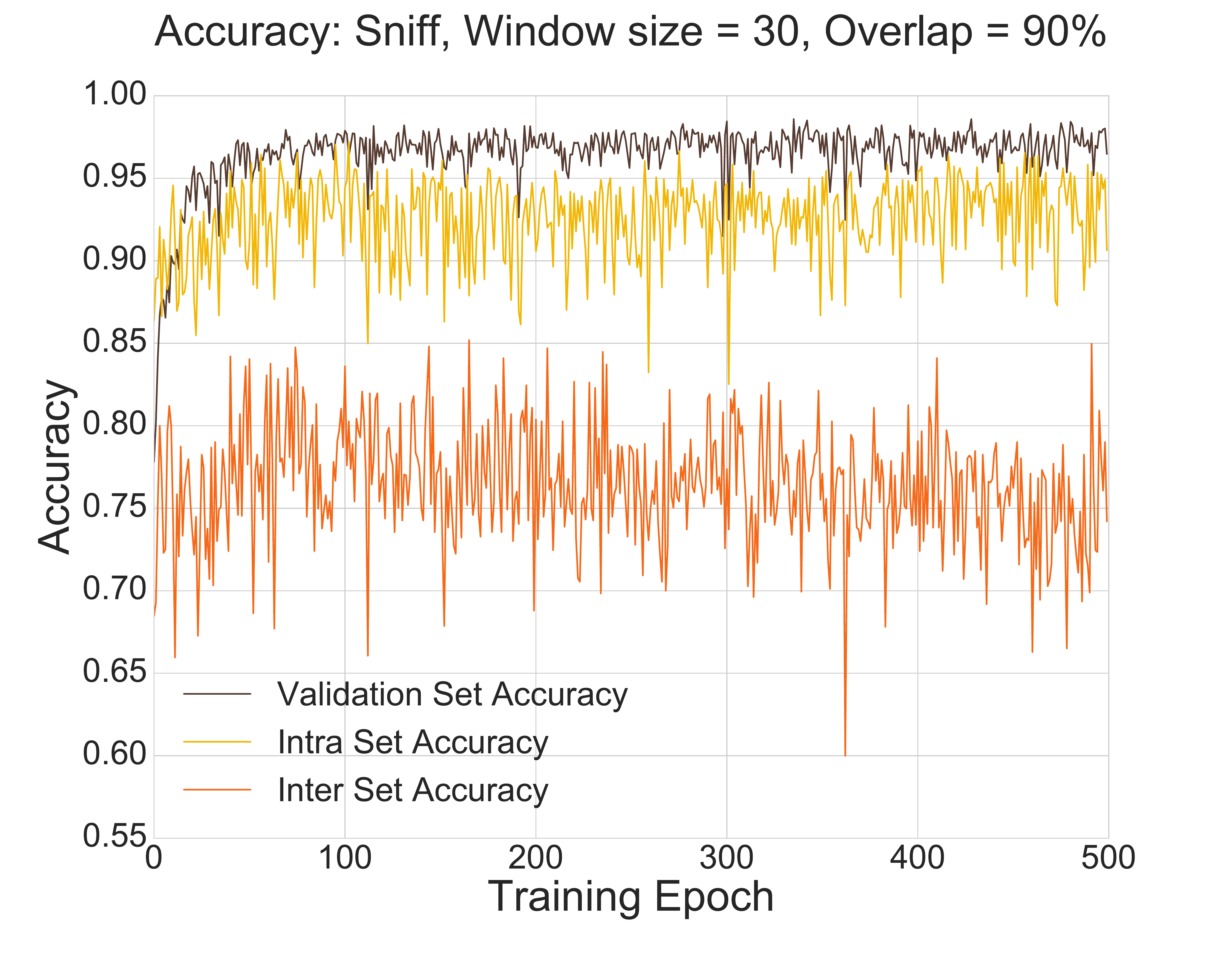}
                 \caption{Accuracy progress}
                 \label{Fig:acc1}
         \end{subfigure}%
         \begin{subfigure}[b]{0.25\textwidth}
                 \centering
                 \includegraphics[width=0.98\textwidth]{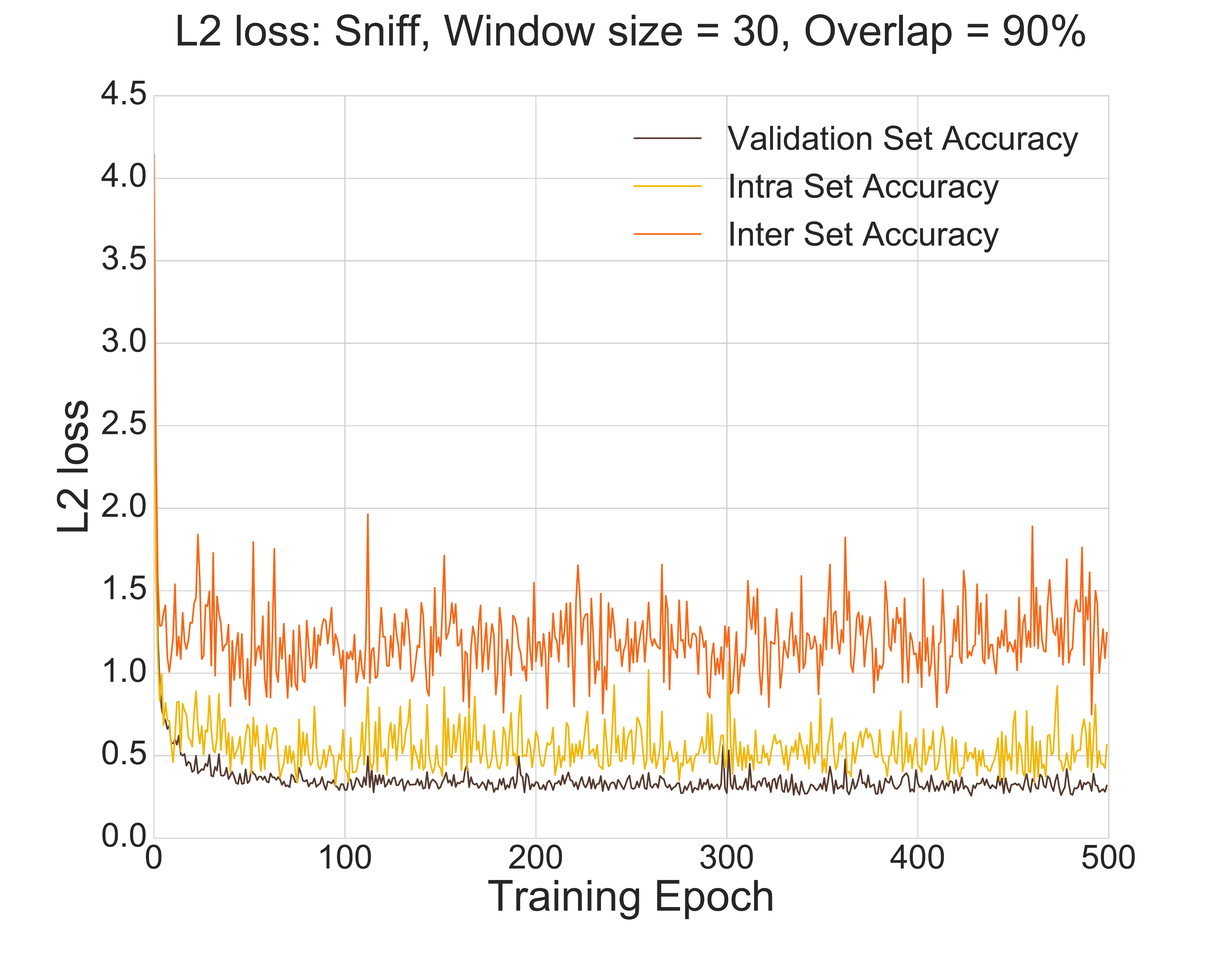}
                 \caption{L2 loss improvement}
                 \label{Fig:acc2}
         \end{subfigure}%

\caption{Training Progress of the RNN over iterations} \label{Fig:devotherappssim}
\end{figure}

We used \emph{early termination} to select the best model. This was because we observed after some iterations the model accuracy reaches to the maximum and stays approximately in the same region for the {\bf \emph{validation set}} whilst the accuracy in {\bf \emph{intra set}} and {\bf \emph{inter set}} shows a slight declining trend as shown in Figure~\ref{Fig:acc1}.  Figure~\ref{Fig:acc2} shows how the L2 loss improved over the training iterations.


To approximately select the \emph{elbow point} of the accuracy graphs, we first applied 20-point moving averaging to the validation set accuracy graph and then selected the point (\emph{elbow point}) from which the accuracy does not improve by 5\% for next four consecutive points. The moving average window and the improvement threshold was decided empirically. Once the elbow point is decided we selected 11 models (five previous models and the five next models including the model at the elbow point). To present the performance results in the next section we selected the models that gave the highest average accuracy over all three datasets. The window and overlap configurations that gave the highest accuracy were: (a) for sniffing, a window size=30, with 90\% overlap; and (b) for deep breathing, a window size=250, also with 90\%. For SVM, we picked the model providing the best cross-validation accuracy. 

\textbf{Model Reduction:} To empirically study the computational overhead vs. latency/accuracy tradeoffs, we compressed the selected RNN models using the in-built 256--level quantization function \cite{quant} provided in Tensorflow. This function extracts the minimum and maximum for each layer, and then compresses each float value to an eight-bit integer. Note that this option will save space in zipped formats that are usually used inside Android applications. We executed the original as well as quantized models.

\section{Results}
\label{Sec:Results}


We utilize four performance metrics for our experimental evaluation: \\ \vspace{-3mm}

\noindent \textbf{i) Accuracy}: The percentage  of correct user identifications.  \\ \vspace{-3mm}

\noindent \textbf{ii) Feature Extraction Time}: Time taken to extract MFCC features from an audio file.   \\ \vspace{-3mm}

\noindent \textbf{iii) Model Loading  Time}: Time to load the machine learning model into the memory.   \\ \vspace{-3mm}

\noindent \textbf{iv) Inference Time}: Time to predict the user label once the feature extraction has been done and learning model is loaded to memory. \\ \vspace{-3mm}

\begin{figure}[t!p]
         \begin{subfigure}[b]{0.25\textwidth}
                 \centering
                 \includegraphics[width=0.98\textwidth]{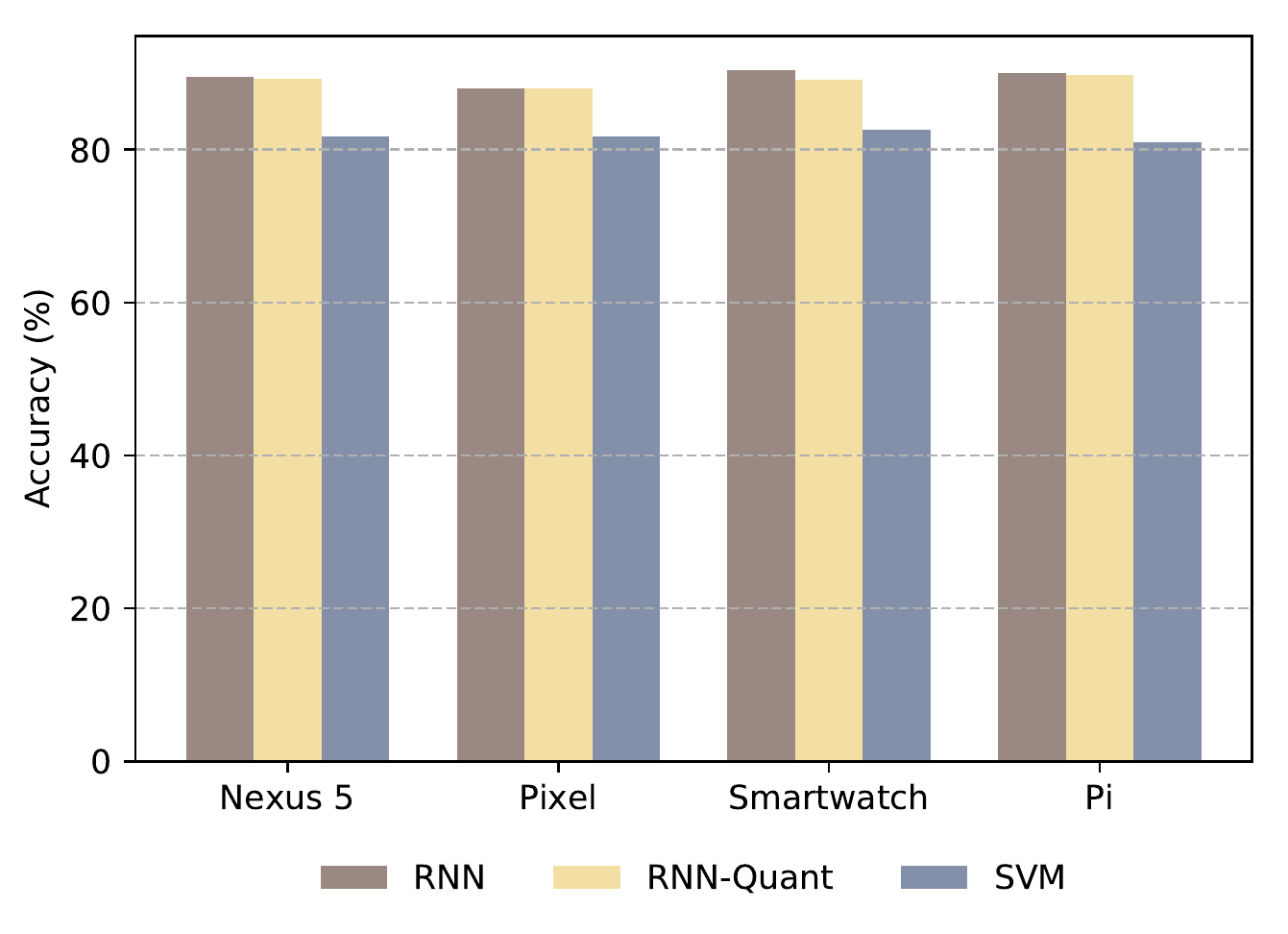}
                 \caption{Accuracy - Sniff}
                 \label{fig2:coolcat}
         \end{subfigure}%
         \begin{subfigure}[b]{0.25\textwidth}
                 \centering
                 \includegraphics[width=0.98\textwidth]{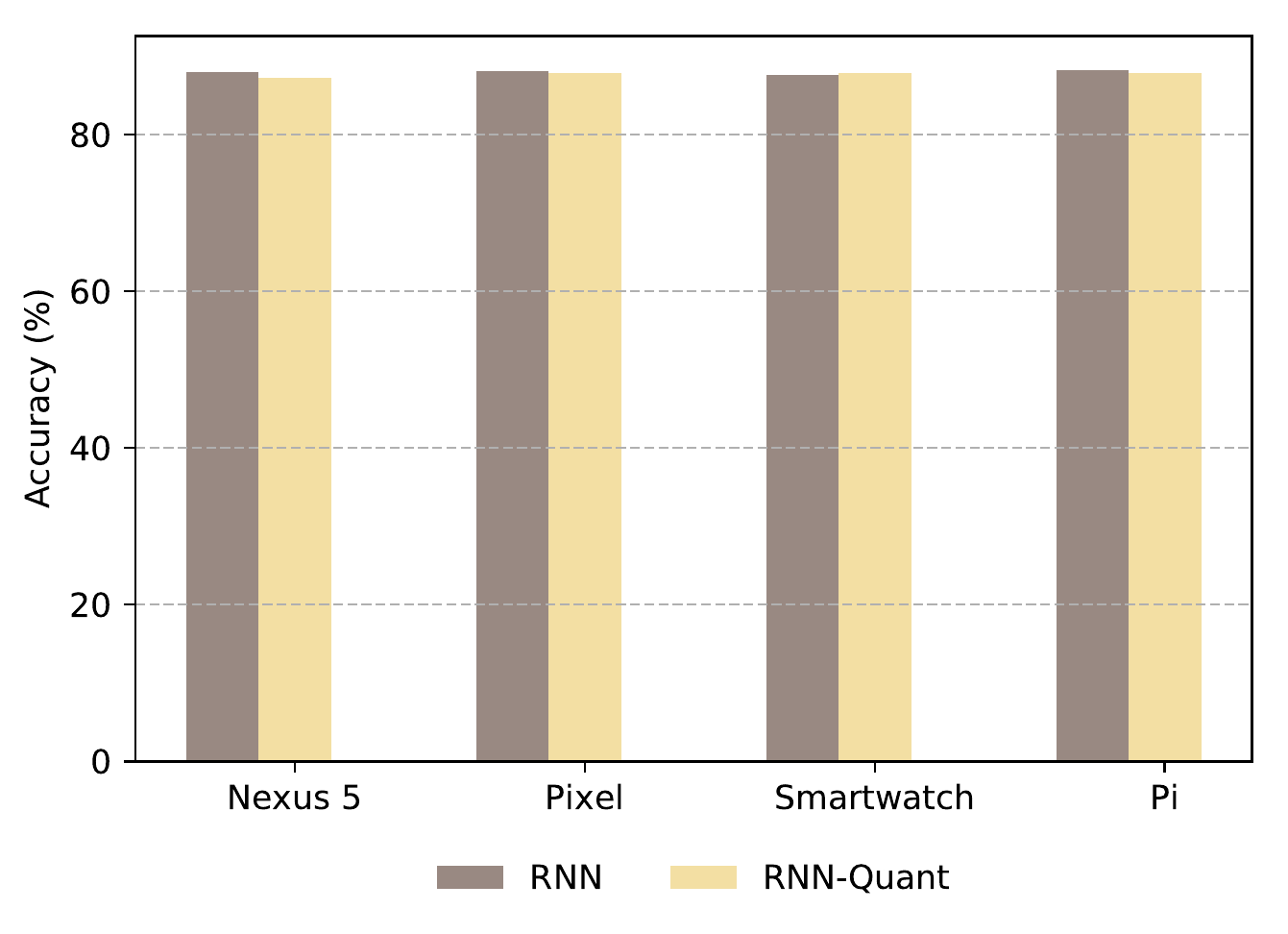}
                 \caption{Accuracy - Deep}
                 \label{fig2:bossycat}
         \end{subfigure}%
         \\
         
         \begin{subfigure}[b]{0.25\textwidth}
                 \centering
                 \includegraphics[width=0.98\textwidth]{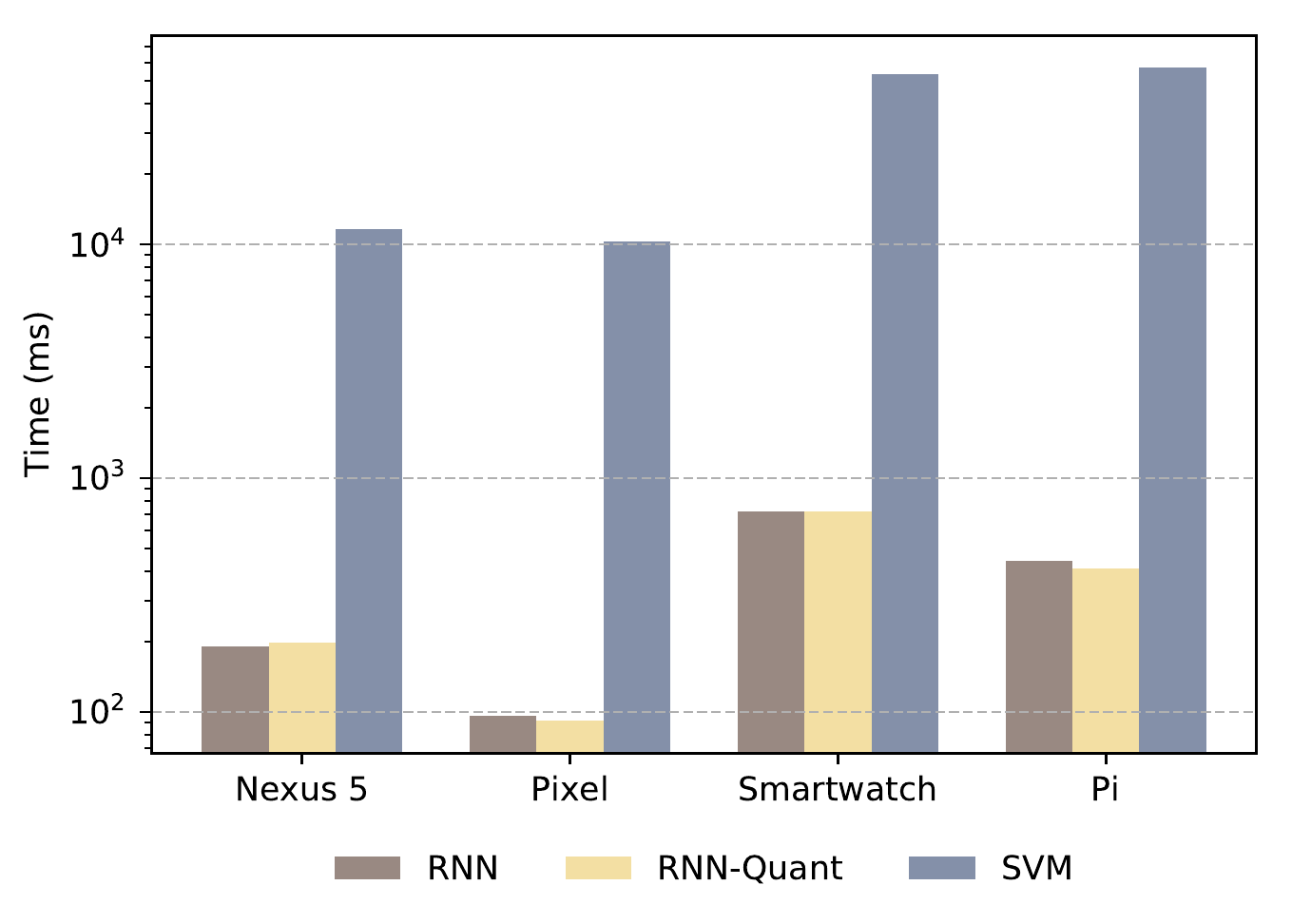}
                 \caption{Model Loading Time - Sniff}
                 \label{fig2:frowningcat}
         \end{subfigure}%
         \begin{subfigure}[b]{0.25\textwidth}
                 \centering
                 \includegraphics[width=0.98\textwidth]{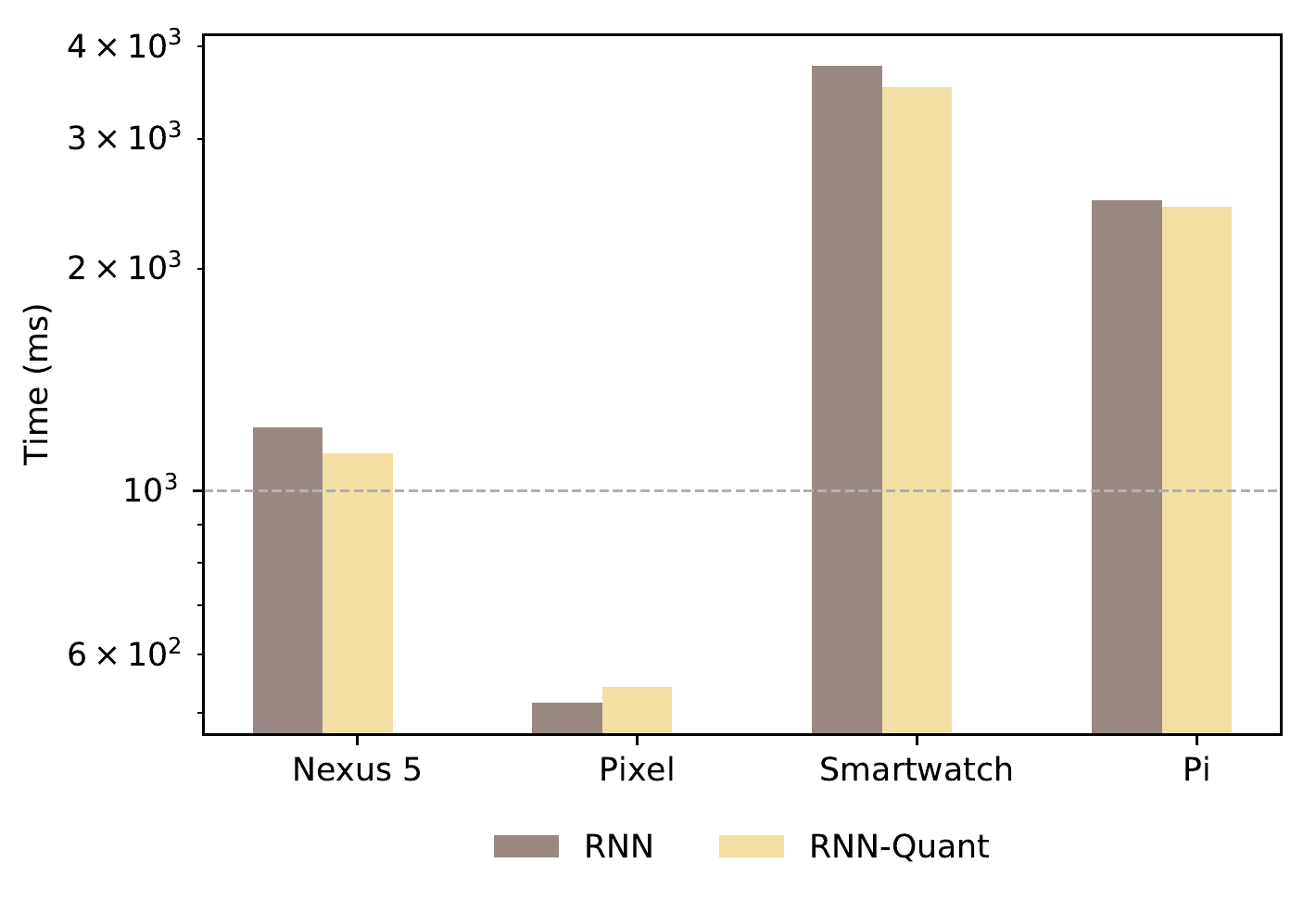}
                 \caption{Model Loading Time - Deep}
                 \label{fig2:bossycat}
         \end{subfigure}%
         \\
         
         \begin{subfigure}[b]{0.25\textwidth}
                 \centering
                 \includegraphics[width=0.98\textwidth]{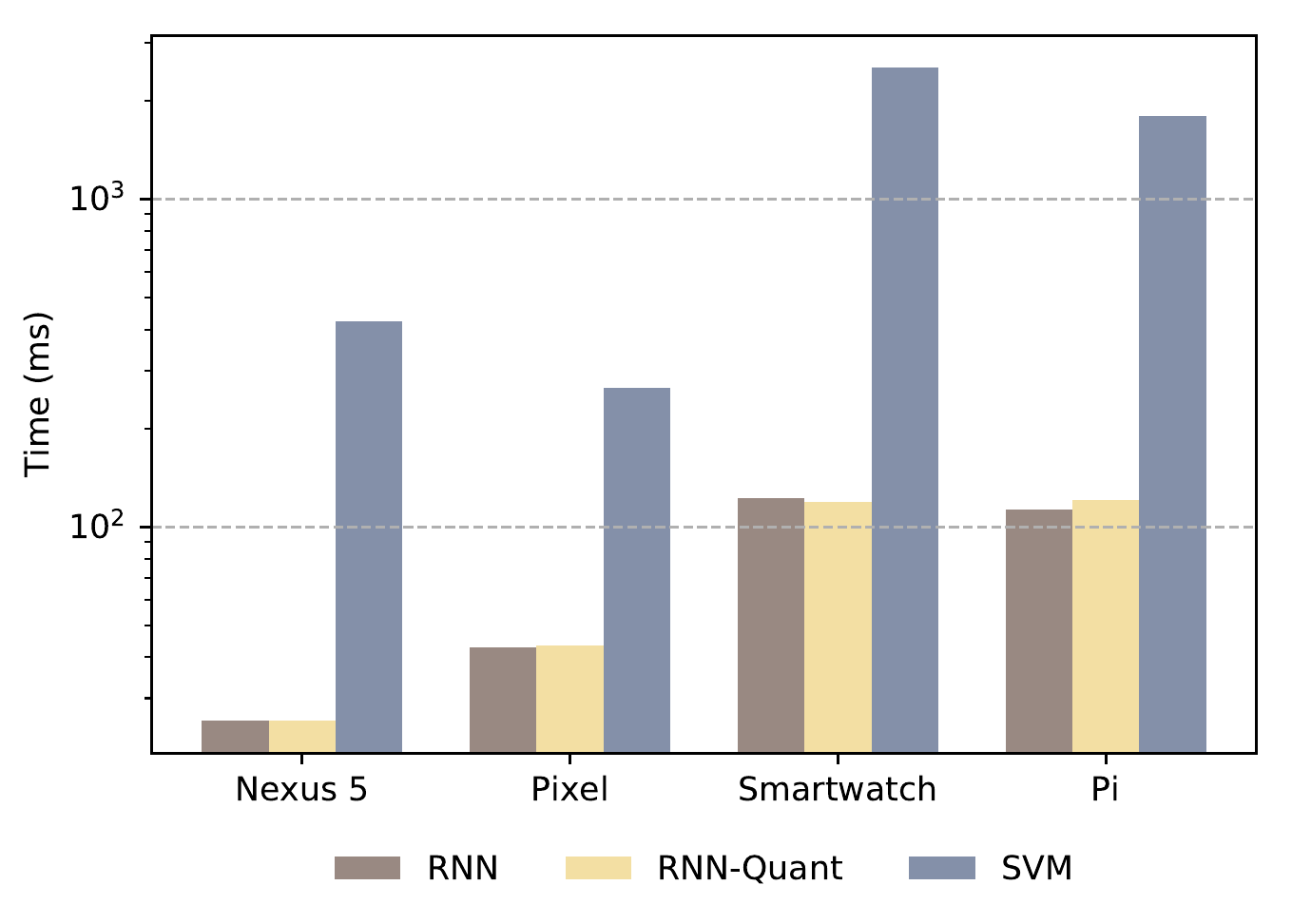}
                 \caption{Inference Time - Sniff}
                 \label{fig2:frowningcat}
         \end{subfigure}%
         \begin{subfigure}[b]{0.25\textwidth}
                 \centering
                 \includegraphics[width=0.98\textwidth]{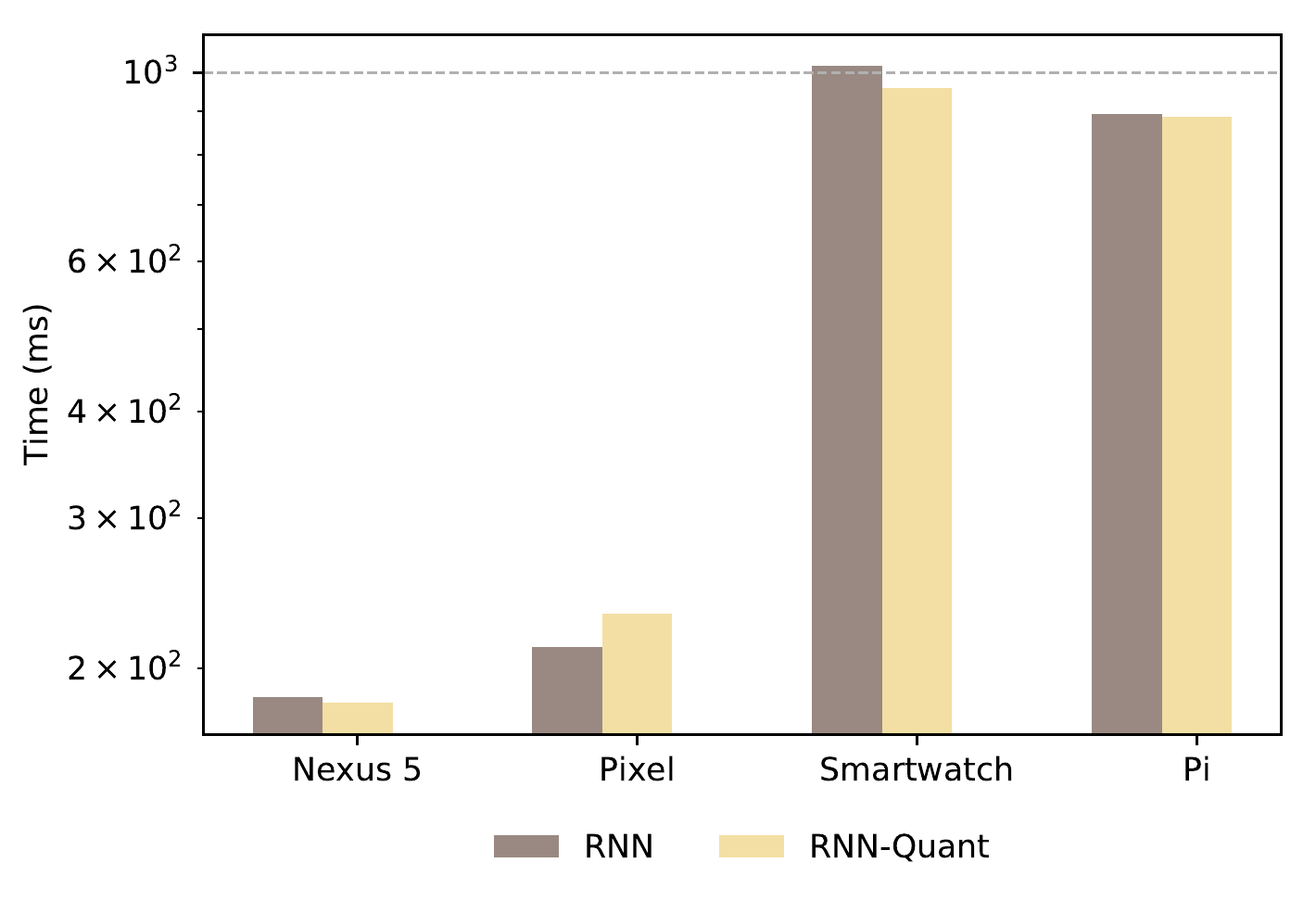}
                 \caption{Inference Time - Deep}
                 \label{fig2:bossycat}
         \end{subfigure}%

\caption{Metrics for Different Breathing Gestures} \label{fig2}
\end{figure}

\subsection{Performance of LSTM}

We report average values of the execution times for different phases of the classification process. The feature extraction time was in the range 40 - 60 ms for sniff, and in the range 126 - 484 ms for deep, across the four devices. As expected, the Pixel (the most powerful platform with the highest memory)  and the smartwatch, impose the least and highest feature extraction times, respectively. Figure~\ref{fig2} plots the results for other three metrics. Note that the time scale on y-axis is in log scale. We can see that:  


\begin{itemize}
\item Model loading time decreases linearly with the RAM of the device.   Loading LSTM model to Pixel takes around 100 ms (4 GB RAM) while the same model takes roughly twice the time (200 ms) on Nexus 5 (2 GB RAM), four times (400 ms) on Pi (1 GB RAM) and seven times (700 ms) on smartwatch (512 MB RAM) for the sniff breathing gesture with LSTM. The same observation holds for the deep breathing gesture. \\ \vspace{-2mm}

\item Inference time depends on the processing power of the devices. The inference time (using the LSTM model) for sniff breathing gesture is on average 23 ms for Nexus 5,  40 ms for Pixel, and around 100 ms for smartwatch and Pi for LSTM model. In contrast, the inference time for deep breathing gesture is approx. 180 ms for Nexus 5, 200 ms for Pixel, 900 ms for Pi and 1000 ms for smartwatch.  \\ \vspace{-2mm}

\item The accuracy of the system  was around 90\% for both deep and sniff breathing gesture for the intra set. We also observe that the accuracy of LSTM models were slightly than SVM. This indicates that the deep learning model can perform at least as well as the alternative SVM approach, even though the volume of training data was fairly small. Note also, that the accuracy of user identification drops to 75\%  and 70\% for sniff and deep breathing gestures, respectively when applied to the \emph{inter set} data (which was excluded entirely from the training set. This result is consistent with our prior results~\cite{chauhan2017breathprint}, and suggests that a larger training corpus will be needed to a wider range of context-dependent variations in breathing patterns. \\ \vspace{-2mm}

\end{itemize}

\subsection{Benefit of Quantized Operation}
Our results (Figure~\ref{fig2} (a) and (b)) also show that the use of a quantized model does not result in any loss of classification accuracy (compared to the full LSTM model). However, quantization reduces the size of the model and offers significant execution benefits. The size of the quantized model is 175 KB (compared to 1.1 MB  for the unoptimized model) for sniff breathing gesture, and 264 KB (compared to 1.1 MB for the unoptimized model) for the deep breathing gesture.  The quantized model thus reduces the memory footprint significantly (by a factor of 4-6), enabling it to be loaded significantly faster as well. \\ \vspace{-2mm} 

\subsection{LSTM vs. SVM}
Contrary to popular belief, the shallow SVM model does provide accuracy comparable to the LSTM model, but takes much longer (10 - 50  seconds) to load into memory.  Moreover, the execution time taken to predict the user label is also 5 - 20 times higher than the corresponding time for the LSTM-based model, across all the devices. The best model we tested with SVM for sniff  breathing gesture is 280 MB in size. Comparatively, the LSTM model size is 2 MB when unoptimized and only a few hundred  KBs when quantized. The large size of SVM models is due to the large number of support vectors needed to support the multi class classification, which does not happen for binary class classification. In fact, the size of the SVM model for deep breathing is 2 GB or higher, and hence could not be loaded onto any of our devices. Our results suggest that the LSTM-based deep model is, in fact, more lightweight and robust (especially when we utilize the quantized LSTM model) than the SVM-based shallow classifier. \\ \vspace{-2mm}

\section{Conclusion}
\label{Sec:Conclusion}

Our experiments reveal that an RNN-based approach for user authentication based on breathing acoustics is not only robust, but is also lightweight enough to be effectively executed on a variety of resource-constrained embedded devices. In particular, an appropriately quantized, LSTM-based deep learning model can authenticate users (using just `deep' and `sniff' breathing gestures) with accuracies higher than 90\%, and utilizes models that are modestly sized (a couple of hundred KB). The resulting user authentication latency is not only small ($<=$200 ms) for representative smartphones, but is within acceptable bounds ($<=$ 1 second) even for the highly resource-limited smartwatch platform. Note that these performance numbers are achieved using CPU-only computation, and should be significantly improved using GPU-offloading approaches proposed by other researchers. Our investigations suggest that RNNs offer a compelling lightweight alternative to CNNs for many sensor-driven pervasive applications, especially if the application utilizes temporal features of the underlying sensor data.

\bibliographystyle{IEEEtran} 
\bibliography{IEEEabrv,biblio}
 
\end{document}